\documentstyle[prl,aps,preprint,epsf,floats]{revtex}

\newcommand{\nwc}{\newcommand}
\nwc{\cl}  {\clubsuit}
\nwc{\di}  {\diamondsuit}
\nwc{\sps} {\spadesuit}
\nwc{\hyp} {\hyphenation}
\nwc{\be}  {\begin{equation}}
\nwc{\ee}  {\end{equation}}
\nwc{\ba}  {\begin{array}}
\nwc{\ea}  {\end{array}}
\nwc{\bdm} {\begin{displaymath}}
\nwc{\edm} {\end{displaymath}}
\nwc{\bea} {\be\ba{rcl}}
\nwc{\eea} {\ea\ee}
\nwc{\ben} {\begin{eqnarray}}
\nwc{\een} {\end{eqnarray}}
\nwc{\bda} {\bdm\ba{lcl}}
\nwc{\eda} {\ea\edm}
\nwc{\bc}  {\begin{center}}
\nwc{\ec}  {\end{center}}
\nwc{\ds}  {\displaystyle}
\nwc{\bmat}{\left(\ba}
\nwc{\emat}{\ea\right)}
\nwc{\non} {\nonumber}
\nwc{\bib} {\bibitem}
\nwc{\lra} {\longrightarrow}
\nwc{\Llra}{\Longleftrightarrow}
\nwc{\ra}  {\rightarrow}
\nwc{\Ra}  {\Rightarrow}
\nwc{\lmt} {\longmapsto}
\nwc{\pa} {\partial}
\nwc{\iy}  {\infty}
\nwc{\ovl}  {\overline}
\nwc{\hm}  {\hspace{3mm}}
\nwc{\lf}  {\left}
\nwc{\ri}  {\right}
\nwc{\lm}  {\limits}
\nwc{\lb}  {\lbrack}
\nwc{\rb}  {\rbrack}
\nwc{\ov}  {\over}
\nwc{\pr}  {\prime}
\nwc{\nnn} {\nonumber \vspace{.2cm} \\ }
\nwc{\Sc}  {{\cal S}}
\nwc{\Lc}  {{\cal L}}
\nwc{\Rc}  {{\cal R}}
\nwc{\Dc}  {{\cal D}}
\nwc{\Oc}  {{\cal O}}
\nwc{\Cc}  {{\cal C}}
\nwc{\Pc}  {{\cal P}}
\nwc{\Mc}  {{\cal M}}
\nwc{\Ec}  {{\cal E}}
\nwc{\Fc}  {{\cal F}}
\nwc{\Hc}  {{\cal H}}
\nwc{\Kc}  {{\cal K}}
\nwc{\Xc}  {{\cal X}}
\nwc{\Gc}  {{\cal G}}
\nwc{\Zc}  {{\cal Z}}
\nwc{\Nc}  {{\cal N}}
\nwc{\fca} {{\cal f}}
\nwc{\xc}  {{\cal x}}
\nwc{\Ac}  {{\cal A}}
\nwc{\Bc}  {{\cal B}}
\nwc{\Uc}  {{\cal U}}
\nwc{\Vc}  {{\cal V}}
\nwc{\Th} {\Theta}
\nwc{\th} {\theta}
\nwc{\vth} {\vartheta}
\nwc{\eps}{\epsilon}
\nwc{\si} {\sigma}
\nwc{\Gm} {\Gamma}
\nwc{\gm} {\gamma}
\nwc{\bt} {\beta}
\nwc{\La} {\Lambda}
\nwc{\la} {\lambda}
\nwc{\om} {\omega}
\nwc{\Om} {\Omega}
\nwc{\dt} {\delta}
\nwc{\Si} {\Sigma}
\nwc{\Dt} {\Delta}
\nwc{\al} {\alpha}
\nwc{\vph}{\varphi}
\nwc{\zt} {\zeta}
\def\VEV#1{\left\langle #1\right\rangle}

\def\pr#1{#1^\prime}
\def\ltap{\raisebox{-.4ex}{\rlap{$\sim$}} \raisebox{.4ex}{$<$}}
\def\gtap{\raisebox{-.4ex}{\rlap{$\sim$}} \raisebox{.4ex}{$>$}}
\nwc{\Id}  {{\bf 1}}
\nwc{\diag} {{\rm diag}}
\nwc{\inv}  {{\rm inv}}
\nwc{\mod}  {{\rm mod}}
\nwc{\hal} {\frac{1}{2}}
\nwc{\tpi}  {2\pi i}

\def\KK{{\rm I\kern -.2em  K}}
\def\NN{{\rm I\kern -.16em N}}
\def\RR{{\rm I\kern -.2em  R}}
\def\ZZ{Z \kern -.43em Z}
\def\QQ{{\rm \kern .25em
             \vrule height1.4ex depth-.12ex width.06em\kern-.31em Q}}
\def\CC{{\rm \kern .25em
             \vrule height1.4ex depth-.12ex width.06em\kern-.31em C}}
\def\ZZZ{Z\kern -0.31em Z}

\def\MeV {\,{\rm  MeV}}

\def\pr#1{Phys. Rev. {\bf #1}}

\def\PL#1{Phys. Lett.~{\bf #1}}
\def\PR#1{Phys. Rev.~{\bf #1}}

\def\PRL#1{Phys. Rev. Lett.~{\bf #1}}

\def\ZP#1{Z. Phys.~{\bf #1}}

\begin{document}
\setcounter{page}{0}
\def\footnoterule{\kern-3pt \hrule width\hsize \kern3pt}
\tighten

\title{The Chiral Phase Transition at High Baryon Density\\ from
  Nonperturbative Flow Equations\thanks {This work is supported in part by
    funds provided by the U.S.~Department of Energy (D.O.E.) under
    cooperative research agreement \# DE-FC02-94ER40818 and by the Deutsche
    Forschungsgemeinschaft.}}

\author{J{\"u}rgen Berges~\footnote{Email: {\tt Berges@ctp.mit.edu}}}

\address{Center for Theoretical Physics \\ 
  Massachusetts Institute of
  Technology \\ 
  Cambridge, Massachusetts 02139}

\author{Dirk--Uwe Jungnickel~\footnote{Email: {\tt
      D.Jungnickel@thphys.uni-heidelberg.de}} and Christof
  Wetterich~\footnote{Email: {\tt C.Wetterich@thphys.uni-heidelberg.de}}}

\address{Institut f{\"u}r Theoretische Physik\\
  Universit{\"a}t Heidelberg\\
  69120 Heidelberg\\
  {~}}

\date{MIT--CTP--2794, HD--THEP--98--57}
\maketitle

\thispagestyle{empty}

\begin{abstract}
  We investigate the QCD chiral phase transition at high baryon number
  density within the linear quark meson model for two flavors.
  The method we employ is based on an exact renormalization group equation
  for the free energy.  Truncated nonperturbative flow equations are
  derived at nonzero chemical potential and temperature. Whereas the
  renormalization group flow leads to spontaneous chiral symmetry breaking
  in vacuum, we find a chiral symmetry restoring first order transition at
  high density. Combined with previous investigations at nonzero
  temperature, the result implies the presence of a tricritical point with
  long--range correlations in the phase diagram.
\end{abstract}

\vspace*{\fill}


\section{Introduction}

The behavior of QCD at high temperature and baryon density is of
fundamental interest and has applications in cosmology, the
astrophysics of neutron stars and the physics of heavy ion
collisions.  Over the past years, considerable progress has been achieved
in our understanding of high temperature QCD, where simulations on the
lattice and universality arguments played an essential role.  
Recently, nonperturbative flow equations, based on the Wilsonian
formulation of the renormalization group, were derived for an effective
linear quark meson model~\cite{Wet93-2,JW,BJW}.  The results of this
approach account for both, the low temperature chiral perturbation theory
domain of validity as well as the high temperature domain of critical
phenomena for two quark flavors~\cite{BJW}.  The method may help to shed
some light on the remaining pressing questions at high temperature, like
the nature of the phase transition for realistic values of the strange
quark mass.

Our knowledge of the high density properties of strongly interacting matter
is rudimentary so far. There are severe problems to use standard simulation
algorithms at nonzero chemical potential on the lattice because of a
complex fermion determinant.  Different nonperturbative methods as the
Wilsonian ``exact renormalization group'' or Schwinger--Dyson equations
seem to present promising alternatives. It is the purpose of this letter to
pursue the former approach based on an exact nonperturbative flow equation
for a scale dependent effective action $\Gamma_k$~\cite{Wet93-2}, which is
the generating functional of the $1 PI$ Green functions in the presence of
an infrared momentum cutoff $\sim k$.  The renormalization group flow for
the average action $\Gm_k$ interpolates between a given short distance or
classical action $S$ and the standard effective action $\Gm$, which is
obtained by following the flow for $\Gm_k$ to $k=0$, thus removing the
infrared cutoff in the end.
 
QCD at nonzero baryon density is expected to have a rich phase structure
with different possible phase transitions as the density is increased. A
prominent example is the liquid--gas nuclear transition which has been
studied in low energy heavy ion collisions. Large efforts in
ultra--relativistic heavy ion collision experiments focus on a transition
at high temperatures and/or densities into what is generally known as the
quark--gluon plasma -- a new phase of matter in which color is screened
rather than confined and chiral symmetry is restored.  In addition to the
nuclear and quark matter phases a number of possibilities like the
formation of meson condensates or strange quark matter have been
investigated.  Recently, an extensive
discussion has focused on the symmetry of the high density ground state,
where condensates of quark Cooper pairs may break the color symmetry
of QCD spontaneously and lead to a superconducting
phase~\cite{CSC,BR}. 
We will concentrate in this work on the chiral phase
transition at nonzero baryon number density and apply our methods to an
effective linear quark meson model for the low energy degrees of freedom of
QCD\footnote{First calculations indicate that diquark condensates have only little influence on the order parameter associated with chiral symmetry
breaking in the vacuum \cite{BR}.}.  
We note that the methods we discuss are equally suited to treat quark
pair condensation and leave this for future work\footnote{In the context of
  quark pair formation the renormalization of quark operators in
  the vicinity of the Fermi surface has been studied in~\cite{RG}.}.

The linear quark meson model we consider can be viewed as a generalization
of the Nambu--Jona-Lasinio model for QCD~\cite{Klevansky}.  Though these
approaches do not explain the confinement of quarks, they are adequate for
many purposes and their predictions show convincing agreement with the
results of real and numerical experiments. Calculations at nonzero baryon
density are typically based on a mean field approximation and it was
claimed~\cite{Klevansky} that the order of the chiral phase transition is
ambiguous. Recently, the order of this high density transition has raised
considerable interest. One can argue on general grounds~\cite{BR,SB} that
if two--flavor QCD has a second order transition (crossover) at high
temperatures and a first order transition at high densities, then there
exists a tricritical point (critical endpoint) with long--range
correlations in the phase diagram. The physics around this point is
governed by universality and may allow for distinctive signatures in heavy
ion collisions~\cite{BR,SB,SRS}.

In this work it is our main concern to show that in a proper treatment
beyond mean field theory the order of the chiral phase
transition can be fixed within the models under investigation. For this
purpose, a crucial observation is the strong attraction of the flow to
partial infrared fixed points~\cite{JW,BJW}. The two remaining relevant or
marginal parameters can be fixed by the phenomenological values of
$f_{\pi}$ and the constituent quark mass. For two massless quark flavors we
find that chiral symmetry restoration occurs via a first order transition
between a phase with low baryon density and a high density phase.

We emphasize, nevertheless, that the linear quark meson model captures the
low density properties of QCD only incompletely since the effects of
confinement are not included.  In particular, for a discussion of the
liquid--gas nuclear transition the inclusion of nucleon degrees of freedom
seems mandatory~\cite{NucMat}. On the other hand this model is expected to
provide a reasonable description of the high density properties of QCD.

\section{Linear Quark Meson Model \label{model}}

It is an important property of the average action $\Gamma_k$ that at any
scale $k$ only fluctuations with momenta in a small range around $k$
influence its flow with $k$.  This implies that for each range of $k$ only
those degrees of freedom have to be included which are relevant around $k$.
In QCD these may comprise compounds of quarks and gluons.  As
long as one concentrates only on the chiral properties of QCD one may
employ for scales $k$ below a ``compositeness scale'' $k_{\Phi}$ a
description in terms of quark and (pseudo--)scalar meson degrees of
freedom~\cite{MG,JW,BJW}. All other degrees of freedom are assumed to be
integrated out.  The scalar and pseudoscalar mesons are described by a
complex field $\Phi$ which transforms as $(\ovl{\bf N},\bf{N})$ under the
chiral flavor group $SU_L(N)\times SU_R(N)$ with $N$ the number of light
quark flavors.  Chiral symmetry breaking to a vector--like subgroup
$SU_V(N)$ occurs through a nonvanishing chiral condensate
$\VEV{\Phi^{ab}}=\ovl{\si}_0\dt^{ab}$.

We will restrict our discussion here to two massless quarks. A
more realistic treatment would have to include the small up and down quark
masses and the finite, but much heavier, strange quark mass. Though their
inclusion will change certain quantitative estimates, they most likely do
not change the qualitative outcome of the investigation discussed below.
In the vacuum the scalar iso--triplet as well as the pseudoscalar
singlet (associated with the $\eta^\prime$ meson) contained in $\Phi$ are
significantly heavier than the pions and the sigma meson. This property
even holds in the symmetric phase where the pions are no longer Goldstone
bosons. We take this as a guide also at nonzero density and
decouple these degrees of freedom. For $N=2$ this can be achieved in a
chirally invariant way.  
This results in a parameterization of $\Phi$ in terms of
the $\si$--meson field and the three pion field components $\pi^l$ as
\begin{equation}
  \label{AAA400}
  \Phi=\frac{1}{2}\left(\si+i\pi^l\tau_l\right)
\end{equation}
where $\tau_l$ $(l=1,2,3)$ denote the Pauli matrices. 

Closed nonperturbative flow equations follow from the exact renormalization
group equation by a suitable truncation of the effective average action. We
consider here a rather simple truncation of the momentum dependence in
terms of standard kinetic terms parameterized by a running meson wave
function renormalization constant $Z_{\Phi,k}$ and a scale dependent Yukawa
coupling $\ovl{h}_k$. For the meson interaction we consider the most
general form of the scalar potential $U_k$ consistent with the symmetries.
For two flavors chiral symmetry implies that $U_k$ depends on
only one invariant $\ovl{\rho}={\rm Tr}(\Phi^{\dagger} \Phi)$.  The
effective potential $U=\lim_{k\to 0}U_k=\lim_{k\to 0}\Gamma_k \, T/V$ is
the relevant quantity for a study of the spontaneous breaking of chiral
symmetry and is obtained for constant field configurations. At its minimum
$\ovl{\rho}_0$ the effective potential is related to the energy density
$\eps$, the entropy density $s$, the quark number density $n_q$ and the
pressure $p$ by
\begin{equation}
  \label{thermodyn}
  U(\ovl{\rho}_0;\mu,T)=\eps-Ts-\mu n_q=-p \; .
\end{equation}
Our truncation precludes any nontrivial momentum dependence of the
effective quark and meson propagators or the interactions.  In particular,
it clearly does not account for confinement effects.  At nonzero
temperature $T$ and chemical potential $\mu$ associated to the
conserved quark number our ansatz for $\Gamma_k$ reads
\begin{eqnarray}
  \label{truncation} 
  \Gm_k &=& \ds{
    \int^{1/T}_0 dx^0\int d^3x \Bigg\{  
    i \ovl{\psi}^a (\gamma^{\mu}\pa_{\mu} + \mu \gamma^0) \psi_a
    +\ovl{h}_k {\ovl{\psi}}^a \left[ \frac{1+\gamma^5}{2} {\Phi_a}^b
    - \frac{1-\gamma^5}{2} {(\Phi^{\dagger})_a}^b\right] \psi_b
    }\nnn 
  && \ds{
    \qquad\qquad\qquad \quad +Z_{\Phi,k} 
    \pa_{\mu}\Phi^*_{ab}\pa^{\mu}\Phi^{ab}
    +U_k(\ovl{\rho};\mu,T)\Bigg\} 
    }\, .
\end{eqnarray}
We note that in the Euclidean formalism nonzero temperature results in
(anti--)periodic boundary conditions for (fermionic) bosonic fields in the
Euclidean time direction with periodicity $1/T$.  A nonzero chemical
potential $\mu$ to lowest order results in the term $\sim i\mu
\ovl{\psi}^a\gamma^0\psi_a$ appearing on the right hand side
of~(\ref{truncation}).  Our approximation neglects the dependence of
$Z_{\Phi,k}$ and $\ovl{h}_k$ on $\mu$ and $T$. We also neglect a possible
difference in normalization of the quark kinetic term and the baryon number
current.  The form of the effective action at the compositeness scale,
$\Gamma_{k_\Phi}[\psi,\Phi]$, serves as an initial value for the
renormalization group flow of $\Gamma_{k}[\psi,\Phi]$ for $k<k_\Phi$.  We
will consider here the case that $Z_{\Phi,k_\Phi}\ll1$.  
The limiting case $Z_{\Phi,k_{\Phi}}=0$ can be considered as a
solution of the corresponding Nambu--Jona-Lasinio model where the effective
four--fermion interaction has been eliminated by the introduction of
auxiliary meson fields.

\section{Flow equation for the effective potential}

The dependence on the infrared cutoff scale $k$ of the effective action
$\Gamma_k$ is given by an exact flow equation~\cite{Wet93-2,Wet90-1}, which
for fermionic fields $\psi$ (quarks) and bosonic fields $\Phi$ (mesons)
reads
\begin{equation}
  \label{frame}
  \frac{\pa}{\pa k}\Gm_k[\psi,\Phi] = \frac{1}{2}{\rm Tr} 
  \left\{ \frac{\pa R_{kB}}
    {\pa k} \left(\Gm^{(2)}_k[\psi,\Phi]+R_k\right)^{-1}  \right\} 
    -{\rm Tr} \left\{ \frac{\pa R_{kF}}
    {\pa k} \left(\Gm^{(2)}_k[\psi,\Phi]+R_k\right)^{-1} 
  \right\} \, .
\end{equation}
Here $\Gamma_k^{(2)}$ is the matrix of second functional derivatives of
$\Gamma_k$ with respect to both fermionic and bosonic field components. The
first trace on the right hand side of~(\ref{frame}) effectively runs only
over the bosonic degrees of freedom. It implies a momentum integration and
a summation over flavor indices. The second trace runs over the
fermionic degrees of freedom and contains in addition a summation over
Dirac and color indices. The exact flow equation closely
resembles a one--loop equation with the difference that the full inverse
propagator $\Gm^{(2)}_k$ at the scale $k$ appears instead of the classical
propagator. The infrared cutoff function $R_k$ has a block substructure
with entries $R_{kB}$ and $R_{kF}$. Here $R_{kB}$ denotes the infrared
cutoff function for the bosonic fields and we employ an exponential cutoff
function
\begin{equation}
R_{kB}=\frac{Z_{\Phi,k}q^2}{{\rm exp}(q^2/k^2)-1} \, . \label{bosir}
\end{equation}
With this choice, $R_{kB}$ acts as an additional mass term $R_{kB}\simeq
Z_{\Phi,k} k^2$ for the low momentum $q^2 \ll k^2$ modes. The infrared
cutoff function for the fermions $R_{kF}$ should be consistent with 
chiral symmetries. This can be
achieved if $R_{kF}$ has the same Lorentz structure as the
kinetic term for free fermions~\cite{Wet90-1}.  In presence of a chemical
potential $\mu$ we use
\begin{equation}
  \label{fermir}
  R_{kF}=(\gamma^\mu q_\mu + i \mu \gamma^0) r_{kF} \, . 
\end{equation}
The effective squared inverse fermionic propagator is then of
the form
\begin{eqnarray}
  \label{fermprop}
  P_{kF} &=& \ds{
    [(q_0+i\mu)^2+\vec{q}^{\,2}](1+r_{kF})^2
    } \nnn 
  &=& \ds{
    (q_0+i \mu)^2+\vec{q}^{\,2}+k^2 \Theta 
    (k_{\Phi}^2-(q_0+i\mu)^2-\vec{q}^{\,2})
    }\; ,
\end{eqnarray}
where the second line defines $r_{kF}$ and one observes that
the fermionic infrared cutoff acts as an additional mass--like
term\footnote{The exponential form (\ref{bosir}) of the cutoff
  function $R_{kB}$ renders the first term on the right hand side of
  (\ref{frame}) both infrared and ultraviolet finite. No need for an
  additional ultraviolet regularization arises in this case. This is
  replaced by the necessary specification of an initial value
  $\Gamma_{k_{\Phi}}$ at the scale $k_{\Phi}$.  Here $k_{\Phi}$ is
  associated with a physical ultraviolet cutoff in the sense that
  effectively all fluctuations with $q^2>k_{\Phi}^2$ are already included
  in $\Gamma_{k_{\Phi}}$. A similar property for the fermionic contribution
  is achieved by the $\Theta$--function in~(\ref{fermprop}).} $\sim k^2$
Here the $\Theta$--function can be
thought of as the limit of some suitably regularized function, e.g.\ 
$\Theta^{\eps}=[{\rm exp}\{(q_0+i \mu)^2+\vec{q}^{\,2}-
k_{\Phi}^2\}/\eps+1]^{-1}$.
 
We compute the flow equation for the effective potential $U_k$ from
equation (\ref{frame}) using the ansatz (\ref{truncation}) for $\Gamma_k$
and we introduce a renormalized field $\rho=Z_{\Phi,k}\ovl{\rho}$ and
Yukawa coupling $h_k=Z_{\Phi,k}^{-1/2}\ovl{h}_k$. The flow equation for
$U_k$
\begin{equation}
  \frac{\pa }{\pa k}U_k(\rho;T,\mu)=\frac{\pa }{\pa k}U_{kB}(\rho;T,\mu)
  +\frac{\pa }{\pa k}U_{kF}(\rho;T,\mu) \label{dtu} \, 
\end{equation}
obtains contributions from  bosonic and fermionic fluctuations,
respectively,
\begin{eqnarray}
  \ds{\frac{\pa }{\pa k}U_{kB}(\rho;T,\mu)} &=& \ds{
    \frac{1}{2} T \sum\limits_n
    \int\limits_{-\infty}^{\infty}
    \frac{d^3\vec{q}}{(2 \pi)^3} \frac{1}{Z_{\Phi,k}}
    \frac{\pa R_{kB}(q^2)}{\pa k} \left\{
    \frac{3}{P_{kB}\left(q^2\right) + U_k'(\rho;T,\mu)}\right. 
    }\nnn 
  && \ds{
    +\left. \frac{1}{P_{kB}\left(q^2\right) + U_k'(\rho;T,\mu)
      + 2 \rho U_k''(\rho;T,\mu)}\right\}  
    \label{dtub} 
    }\, ,\\[2mm]
  \ds{\frac{\pa }{\pa k}U_{kF}(\rho;T,\mu)} &=& \ds{
    -8 N_c T \sum\limits_n
    \int\limits_{-\infty}^{\infty}
    \frac{d^3\vec{q}}{(2\pi)^3} 
    \frac{ k \, \Theta (k_{\Phi}^2-(q_0+i \mu)^2-\vec{q}^{\,2})}
    {P_{kF}\left((q_0+i\mu)^2+\vec{q}^{\,2}\right)
      +h_k^2 \rho/2} 
    \label{dtuf} 
    }\, .
\end{eqnarray}
Here $q^2=q_0^2+\vec{q}^{\,2}$
with $q_0(n)=2n\pi T$ for bosons, $q_0(n)=(2n+1) \pi T$ for fermions $(n
\in \ZZ)$ and $N_c=3$ denotes the number of colors.  The scale dependent
propagators on the right hand side contain the momentum dependent pieces
$P_{kB}=q^2+Z_{\Phi,k}^{-1}R_k(q^2)$ and $P_{kF}$ given by~(\ref{fermprop})
as well as mass terms, where $U_k'$, $U_k''$ denote the first and second
derivative of the potential with respect to $\rho$.  The only explicit
dependence on the chemical potential $\mu$ appears in the fermionic
contribution (\ref{dtuf}) to the flow equation for $U_k$. It is instructive
to perform the summation of the Matsubara modes explicitly for the
fermionic part. Since the flow equations only involve one momentum
integration, standard techniques for one loop expressions apply and we find
\begin{eqnarray}
  \lefteqn{
    \frac{\pa }{\pa k}U_{kF}(\rho;T,\mu) = -8 N_c 
    \int\limits_{-\infty}^{\infty}\,\,\,
    \frac{d^4q}{(2 \pi)^4} \frac{k \, \Theta (k_{\Phi}^2-q^2)}
    {q^2+k^2  + h_k^2 \rho/2}  
    +4 N_c \int\limits_{-\infty}^{\infty}
    \frac{d^3\vec{q}}{(2 \pi)^3} 
    \frac{k}{\sqrt{\vec{q}^{\,2}+k^2+h_k^2 \rho/2}} 
    }\nnn && \ds{
    \quad \times
    \left\{ \frac{1}
      {\ds \exp\left[(\sqrt{\vec{q}^{\,2}+k^2+h_k^2 \rho/2}-\mu)/T\right]+1}
      +\frac{1}
      {\ds \exp\left[(\sqrt{\vec{q}^{\,2}+k^2+h_k^2 \rho/2}+\mu)/T\right]+1}
      \label{uapprox}
    \right\} }
\end{eqnarray}
where, for simplicity, we sent $k_{\Phi}\to\infty$ in the $\mu,T$ dependent
second integral. This is justified by the fact that in the
$\mu,T$ dependent part the high momentum modes are exponentially
suppressed.

For comparison, we note that within the present approach one obtains
standard mean field theory results for the free energy if the
meson fluctuations are neglected, $\partial U_{kB}/\partial k \equiv 0$,
and the Yukawa coupling is kept constant, $h_k=h$ in (\ref{uapprox}). The
remaining flow equation for the fermionic contribution could then easily be
integrated with the (mean field) initial condition 
$U_{k_{\Phi}}(\rho)=\ovl{m}_{k_\Phi}^2\rho$. In the following we will
concentrate on the case of vanishing temperature\footnote{For a solution of
  the flow equations at nonzero temperature and vanishing chemical
  potential see~\cite{BJW}.}.  We find (see below) that a mean field
treatment yields relatively good estimates only for the
$\mu$--dependent part of the free energy $U(\rho;\mu)-U(\rho;0)$. On
the other hand, mean field theory does not give a very reliable
description of the vacuum properties which are important for a
determination of the order of the phase transition at $\mu \not = 0$.

\section{Renormalization group flow at nonzero chemical potential}

In the limit of vanishing temperature one expects and observes a
non--analytic behavior of the $\mu$--dependent integrand of the fermionic
contribution (\ref{uapprox}) to the flow equation for $U_k$ because of the
formation of Fermi surfaces. Indeed, the explicit $\mu$--dependence of the
flow equation reduces to a step function
\begin{eqnarray}
  \label{dtuf0} 
  \ds{\frac{\pa }{\pa k}U_{kF}(\rho;\mu) =} &-& \ds{
    8 N_c 
    \int\limits_{-\infty}^{\infty}\,\,\,
    \frac{d^4q}{(2 \pi)^4} \frac{k \Theta (k_{\Phi}^2-q^2)}
    {q^2+k^2  + h_k^2 \rho/2}  
    }\nnn 
  &+& \ds{
    4 N_c \int\limits_{-\infty}^{\infty}
    \frac{d^3\vec{q}}{(2 \pi)^3} 
    \frac{k}{\sqrt{\vec{q}^{\,2}+k^2+h_k^2 \rho/2}} \,\,\,\,
    \Theta\! \left( \mu-\sqrt{\vec{q}^{\,2}+k^2+h_k^2 \rho/2} \right) 
    }\, .
\end{eqnarray}
The quark chemical potential $\mu$ enters the bosonic part (\ref{dtub}) of
the flow equation only implicitly through the meson mass terms
$U_k'(\rho;\mu)$ and $U_k'(\rho;\mu) + 2 \rho U_k''(\rho;\mu)$ for the
pions and the $\si$--meson, respectively.  For scales $k > \mu$ the
$\Theta$--function in (\ref{dtuf0}) vanishes identically and there is no
distinction between the vacuum evolution and the $\mu\not = 0$ evolution.
This is due to the fact that our infrared cutoff adds to the effective
quark mass $(k^2+h_k^2 \rho/2)^{1/2}$. For a chemical potential smaller
than this effective mass the ``density'' $-\partial U_k/\partial \mu$
vanishes whereas for larger $\mu$ one can view
$\mu=[\vec{q}_F^{\,2}(\mu,k,\rho)+k^2+h_k^2 \rho/2]^{1/2}$ as an effective
Fermi energy for given $k$ and $\rho$. A small infrared cutoff $k$ removes
the fluctuations with momenta in a shell close to the physical Fermi
surface\footnote{If one neglects the mesonic fluctuations one can
  perform the $k$--integration of the flow equation~(\ref{dtuf0}) in the
  limit of a $k$--independent Yukawa coupling. One recovers (for
  $k_\Phi^2\gg k^2+h^2\rho/2,\mu^2$) mean field theory results except for a
  shift in the mass, $h^2\rho/2\ra h^2\rho/2+k^2$, and the fact that modes
  within a shell of three--momenta
  $\mu^2-h^2\rho/2-k^2\le\vec{q}^{\,2}\le\mu^2-h^2\rho/2$ are not yet
  included. Because of the mass shift the cutoff $k$ also suppresses the
  modes with $q^2<k^2$.}
$\mu^2-h_k^2\rho/2-k^2<q^2<\mu^2-h_{k=0}^2\rho/2$. Our flow equation
realizes the general idea~\cite{Pol92-1} that for $\mu\neq0$ the lowering
of the infrared cutoff $k\ra0$ should correspond to an approach to the
physical Fermi surface. For a computation of the meson effective potential
the approach to the Fermi surface in (\ref{dtuf0}) proceeds from below and
for large $k$ the effects of the Fermi surface are absent.  By lowering $k$
one ``fills the Fermi sea''.

In vacuum the evolution of $U_k$, the Yukawa coupling $h_k$ and the meson
wave function renormalization $Z_{\Phi,k}$ have been computed
earlier~\cite{JW,BJW}\footnote{In~\cite{JW,BJW} a different variant of the
  fermionic infrared cutoff function $R_{kF}$ than~(\ref{fermprop})
  was used.}. The flow equations for $h_k$ and $Z_{\Phi,k}$ can
be derived from (\ref{frame}) using the truncation (\ref{truncation}) and
one finds that for large enough $k$ their running is well approximated
by~\cite{JW,BJW} $(k < k_{\Phi})$
\begin{equation}
  h_k^2 \simeq Z_{\Phi,k}^{-1} \simeq -\frac{8 \pi^2}{N_c \, 
    {\rm ln}(k/k_{\Phi})} 
  \label{dkh} \, .
\end{equation}
One observes a very strong Yukawa coupling $h_k$ for $k$ in the vicinity of
$k_{\Phi}$. (The Landau pole at $k=k_{\Phi}$ corresponds to taking
$Z_{\Phi,k_{\Phi}} \to 0$ .)  This strong Yukawa coupling implies a fast
approach of running couplings to partial infrared fixed
points~\cite{JW,BJW}.  More precisely, for a rescaled potential
$U_k(\rho)/k^4=\sum_{n=0}^{\infty} \frac{1}{n!}u^{(n)}_k (\rho/k^2)^n$ the
running dimensionless couplings $u^{(n)}_k$ for $n \ge 2$ approach values
proportional to powers of the Yukawa coupling $h_k$, with $u^{(2)}_k/h_k^2
\simeq 1$, $u^{(3)}_k/h_k^6 \simeq -0.0164$, $u^{(4)}_k/h_k^8 \simeq
0.0105$ and similarly for the higher couplings~\cite{BJW}.  In consequence,
the details of the short distance meson interactions become unimportant and
besides the marginal coupling $h_k$ the only relevant parameter of the
model corresponds to the mass parameter $u^{(1)}_k$.  We note
that the large value of $h_k$ is phenomenologically suggested by the
comparably large value of the constituent quark mass.
The observed fixed point behavior allows us to fix the model by
only two phenomenological input parameters and we use 
$f_{\pi}=92.4\MeV$ and $300\MeV\ltap M_q\ltap 350\MeV$.
Previous results for the evolution in vacuum~\cite{JW,BJW} show that for
scales not much smaller than $k_{\Phi}\simeq 650 \MeV$ chiral symmetry
remains unbroken. This holds down to a scale $k_{\chi SB} \simeq 450\MeV$
at which the meson potential $U_k(\rho)$ develops a minimum at
$\rho_{0,k}>0$ thus breaking chiral symmetry spontaneously.  Below the
chiral symmetry breaking scale running couplings are no longer governed by
the partial fixed point. In particular, for $k \ltap k_{\chi SB}$ the
Yukawa coupling $h_k$ and the meson wave function renormalization
$Z_{\Phi,k}$ depend only weakly on $k$ and approach their infrared values.

At $\mu \not = 0$ we will follow the evolution from $k=k_{\chi SB}$ to
$k=0$ and neglect the $k$--dependence of $h_k$ and $Z_{\Phi,k}$ in this
range.  According to the above discussion the initial value $U_{k_{\chi
    SB}}$ is $\mu$--independent for $\mu < k_{\chi SB}$. At this scale
$U_{k_{\chi SB}}$ is well approximated by the polynomial form where the
coefficients $u^{(n)}_{k_{\chi SB}}$, $n=2,3,4$, are estimated by their
partial infrared fixed point values and coefficients with $n>4$ are
neglected. The scalar mass parameter $u^{(1)}_{k}$ crosses from positive to
negative values at the chiral symmetry breaking scale and thus vanishes at
$k=k_{\chi SB}$. The constant $u^{(0)}_{k_{\chi SB}}$ is fixed such that
the pressure of the physical vacuum is zero. We do not assume a polynomial
form for $U_k(\rho;\mu)$ for $k < k_{\chi SB}$ and solve the flow equation
(\ref{dtu}) with (\ref{dtub}), (\ref{dtuf0}) numerically\footnote{Nonzero
  current quark masses, which are neglected in our approximation, result in
  a pion mass threshold and would effectively stop the renormalization
  group flow of renormalized couplings at a scale around $m_{\pi}$. To
  mimic this effect one may stop the evolution by hand at $k_f \simeq
  m_{\pi}$.  We observe that our results are very insensitive to such a
  procedure which can be understood from the fact that a small infrared
  cutoff -- induced by $k_f$ or by the nonzero current quark masses --
  plays only a minor role for a sufficiently strong first order transition.
  For the results presented in the next section we use $k_f=100 \MeV$. The
  flow of the potential $U_k$ around its minima and for its outer convex
  part stabilizes already for $k$ somewhat larger than $k_f$. Fluctuations
  on larger length scales lead to a flattening of the barrier between the
  minima. The approach to convexity is not relevant for the present
  discussion.} as a partial differential equation for the potential
depending on the two variables $\rho$ and $k$ for given $\mu$.

In the fermionic part (\ref{dtuf0}) of the flow equation the
vacuum and the $\mu$--dependent term contribute with opposite signs. This
cancelation of quark fluctuations with momenta below the Fermi
surface is crucial for the restoration of chiral symmetry at high
density.  In vacuum, spontaneous chiral symmetry breaking is induced in
our model by quark fluctuations which drive the scalar mass
term $U_k'(\rho=0)$ from positive to negative values at the scale $k =
k_{\chi SB}$.  (Meson fluctuations have the tendency to restore chiral
symmetry because of the opposite relative sign,
cf.~(\ref{frame}).) As the chemical potential becomes larger than the
effective mass $(k^2+h_k^2 \rho/2)^{1/2}$ quark fluctuations with momenta
smaller than $\vec{q}_F^{\,2}(\mu,k,\rho)=\mu^2-k^2-h_k^2 \rho/2$
are suppressed.  Since $\vec{q}_F^{\,2}$ is monotonically
decreasing with $\rho$ for given $\mu$ and $k$ the origin of the effective
potential is particularly affected.  We will see in the next section that
for large enough $\mu$ this leads to a second minimum of
$U_{k=0}(\rho;\mu)$ at $\rho=0$ and a chiral symmetry restoring first order
transition.

\section{High Density Chiral Phase Transition}

\begin{figure}[t]
  \unitlength1.0cm
  \begin{center}
  \begin{picture}(13.0,8.0)
  \put(0.0,0.5){
  \epsfysize=8.cm
  \epsfbox[140 525 525 760]{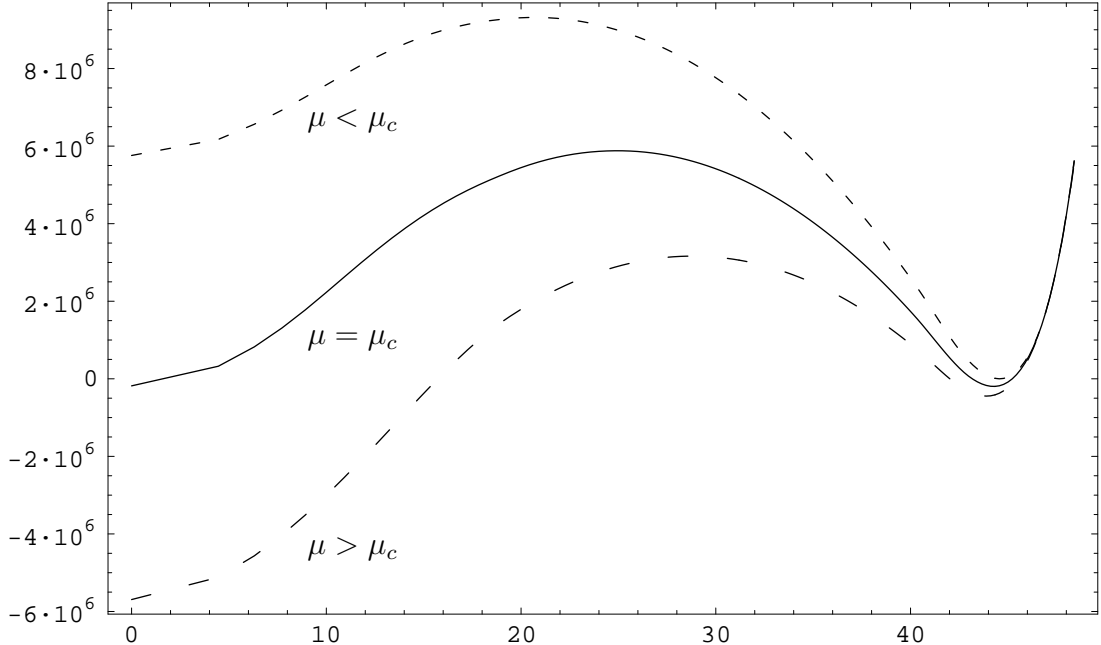}}
  \put(2.8,6.9){$\ds{\mu<\mu_c}$}
  \put(2.8,4){$\ds{\mu=\mu_c}$}
  \put(2.8,1.2){$\ds{\mu>\mu_c}$}
  \end{picture}
  \end{center}
\caption{The zero temperature effective potential $U$ (in ${\rm MeV}^4$)
  as a function of $\si\equiv(\rho/2)^{1/2}$ for different chemical
  potentials. One observes two degenerate minima for a critical chemical
  potential $\mu_c/M_q=1.025$ corresponding to a first order phase
  transition at which two phases have equal pressure and can coexist
  ($M_q=316.2 \MeV$). \label{potps}}
\end{figure}

\begin{figure}[t]
  \unitlength1.0cm
  \begin{center}
  \begin{picture}(13.0,8.0)
  \put(0.3,0.){
  \epsfysize=8.cm
  \epsfbox[125 415 560 700]{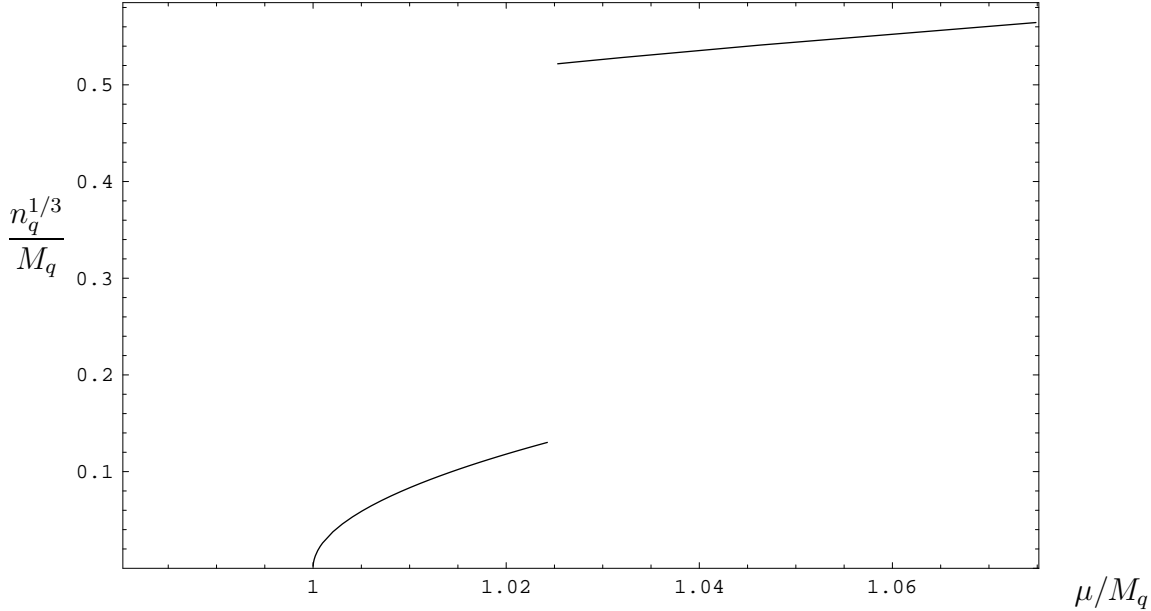}}
  \put(-1.2,5.2){$\ds{\frac{n_q^{1/3}}{M_q}}$}
  \put(13.0,0.5){$\ds{\mu/M_q}$}
  \end{picture}
  \end{center}
\caption{The plot shows $n_q^{1/3}$, where $n_q$ denotes
  the quark number density as a function of $\mu$ in units of the effective
  constituent quark mass ($M_q=316.2\MeV$).} 
  \label{densityps}
\end{figure}

In vacuum or at zero density the effective potential $U(\si)$,
$\si\equiv\sqrt{\rho/2}$, has its minimum at a nonvanishing value
$\si_0=f_{\pi}/2$ corresponding to spontaneously broken chiral symmetry.
As the quark chemical potential $\mu$ increases, $U$ can develop different
local minima. The lowest minimum corresponds to the state of lowest free
energy and is favored.  In Figure~\ref{potps}
we plot the free energy as a function of $\si$ for different values
of the chemical potential $\mu=322.6, 324.0,325.2$ MeV.  Here the effective
constituent quark mass is $M_q=316.2 \MeV$.
We observe that for $\mu < M_q$ the potential at its minimum does not
change with $\mu$.  Since (cf.\ (\ref{thermodyn}))
\begin{equation}
  n_q=-\frac{\pa U}{\pa \mu}_{|{\rm min}}
\end{equation}
we conclude that the corresponding phase has zero density. In contrast,
for a chemical potential larger than $M_q$ we find a low density 
phase where chiral symmetry is still 
broken. The quark number density as a function of $\mu$ is shown in
Figure~\ref{densityps}. One clearly observes the non--analytic behavior
at $\mu=M_q$ which denotes the ``onset'' value for nonzero density. 
From Figure \ref{potps} one also notices the appearance of an additional
local minimum at the origin of $U$.  As the pressure $p=-U$ increases in
the low density phase with increasing $\mu$, a critical value $\mu_c$ is
reached at which there are two degenerate potential minima. Before $\mu$
can increase any further the system undergoes a first order phase
transition at which two phases have equal pressure and can coexist.  In the
high density phase chiral symmetry is restored as can be seen from the
vanishing order parameter for $\mu>\mu_c$. The curvature of the potential
at the origin for $\mu \, \gtap \, \mu_c$ is 
$\partial^2 U/\partial \sigma^2(\sigma=0) \simeq (1.45 M_q)^2$. 
We note that the
relevant scale for the first order transition is $M_q$. 
For this reason we
have scaled our results for dimensionful quantities in units of $M_q$. 

For the class of quark meson models considered here (with $M_q/f_{\pi}$ in
a realistic range around $3$ -- $4$) the first order nature of the high
density transition has been clearly established. In particular, these
models comprise the corresponding Nambu--Jona-Lasinio models where the
effective fermion interaction has been eliminated by the introduction of
auxiliary bosonic fields.  Our treatment, which is based on a
nonperturbative evolution equation for the free energy, goes beyond mean
field theory and puts it into a more systematic context.  For the quark
meson model this method has been used in the past to study the chiral phase
transition at nonzero temperature, including the critical properties near
the second order transition for two massless flavors characteristic of the
three dimensional $O(4)$ universality class~\cite{BJW}.  In the chiral
limit, we therefore find a second order chiral transition at zero chemical
potential and a first order chiral transition at zero temperature. By
continuity these transitions meet at a tricritical point in the
($\mu,T$)--plane~\cite{BR,SB}.  Away from the chiral limit, the second
order chiral transition turns into a smooth crossover. The first order line
of transitions at low temperatures now terminates in a critical endpoint
with long--range correlations~\cite{BR,SB}.  Using the method employed here
a precision estimate of the universal critical equation of state of this
Ising endpoint has been given in~\cite{BTW95}.

The extent to which the transition from nuclear matter to quark matter in
QCD differs from the transition of a quark gas to quark matter in the
quark meson model or NJL--type models has to be clarified by further
investigations. At the phase transition, the quark number density in the
symmetric phase $n_{q,c}^{1/3}=0.52M_q$ turns out in this model to be not
much larger than nuclear matter density, $n_{q,nuc}^{1/3}=152\MeV$.
Here the open problems are related to the description of the low density
phase rather than the high density phase.  Quite generally, the inclusion
of nucleon degrees of freedom for the description of the low density phase
shifts the transition to a larger chemical potential and larger baryon
number density for both the nuclear and the quark matter
phases~\cite{NucMat}. An increase in $\mu_c$ results also from
the inclusion of vector mesons. This would not change the
topology of the phase diagram inferred from our model.  
Nevertheless, the topology of the
QCD phase diagram is also closely connected with the still unsettled
question of the order of the high temperature ($\mu=0$) transition for
three flavors with realistic quark masses. If the strange quark mass is too
small, or if the axial $U(1)$ symmetry is effectively restored in the
vicinity of the transition, then one may have a first order transition at
high $T$ which is driven by fluctuations~\cite{PW}. In this case, critical
endpoints only occur if the two first order regions at $T \gg \mu$ and $\mu
\gg T$ are disconnected.

Finally, the low--density first order transition 
from a gas of nucleons or the
vacuum to nuclear matter (nuclear gas--liquid transition) can only be
understood if the low--momentum fermionic degrees of freedom are described
by nucleons rather that quarks~\cite{NucMat}. The low--density branch of
Figure~\ref{densityps} cannot be carried over to QCD. We emphasize that
nucleon degrees of freedom can be included in the framework of
nonperturbative flow equations.

\acknowledgments

We thank K.\ Rajagopal and D.\ T.\ Son for helpful conversations.

\end{document}